\documentclass[aps,prd,preprint,amsmath,amssymb]{revtex4}
\usepackage{bm}
\usepackage{enumerate}

\def\be{\begin{enumerate}}
\def\ee{\end{enumerate}}
\def\beq{\begin{equation}}
\def\eeq{\end{equation}}
\def\bea{\begin{eqnarray}}
\def\eea{\end{eqnarray}}
\def\a{\alpha}

 \def\half{\textstyle{\frac{1}{2}}}
\def\3halfs{\textstyle{\frac{3}{2}}}

\def\ph{\phantom}

\def\ben{\begin{enumerate}}
\def\een{\end{enumerate}}
\def\bitem{\begin{itemize}}
\def\eitem{\end{itemize}}
\def\half{\frac{1}{2}}

\begin{document}
\title{Post-Newtonian parameters and
constraints on Einstein--aether theory}
\author{Brendan Z. Foster}
\email[]{bzf@umd.edu}
\author{Ted Jacobson}
\email[]{jacobson@umd.edu}

\affiliation{Dept.~of Physics, University of Maryland, College
Park, MD 20742-4111, USA}

\date{February 7, 2006}
%
%
\begin{abstract}
We analyze the observational and theoretical constraints on
 ``Einstein--aether theory", a generally covariant theory
 of gravity coupled to a dynamical,
unit, timelike vector field that breaks local Lorentz symmetry.
The results of a computation of the remaining post-Newtonian
parameters are reported. These are combined with other results to
determine the joint post-Newtonian, vacuum-\v{C}erenkov,
nucleosynthesis, stability, and positive-energy constraints. All
of these constraints are satisfied by parameters in a large
two-dimensional region in the four-dimensional parameter space
defining the theory.
\end{abstract}
\maketitle
%
%
\section*{}
%
%
Alternative theories of gravity that deviate from general
relativity have been ruled out or severely constrained
systematically as observations have improved~\cite{WillBook,
WillReview}. At this stage the most studied surviving alternative
is ``scalar-tensor theory", of which Jordan--Brans--Dicke is a
well-known example. This sort of theory is a simple extension of
general relativity containing a fundamental scalar field. The
existence of such scalar fields has been suggested by moduli
fields that arise in higher dimensional approaches to gravity and
quantum gravity, such as string theory.

Vector-tensor theories consisting of general relativity coupled to
a dynamical vector field are much more complicated than
scalar-tensor theories, due to the metric derivatives that appear
in the covariant derivative of the vector field.
They also appear
to suffer from a glaring problem: because of the indefinite
signature of the spacetime metric, some of the degrees of freedom
are always associated with negative energies.
This problem need not occur, however, if the vector
field is constrained to have a fixed timelike magnitude.
Such a vector field specifies a
particular rest frame at each point of spacetime; hence, it
``spontaneously" breaks the local Lorentz symmetry in a dynamical
fashion. These theories---of which there is a four-parameter
family---preserve the full diffeomorphism
symmetry group of general relativity,
and gravity is still described by the
curvature of the spacetime metric. The
existence of such a vector field is not as well-motivated
as a scalar field, however a number of approaches to
quantum gravity have very tentatively suggested that
Lorentz symmetry might be broken. If general covariance
is to be preserved, then any Lorentz violating vector (or tensor)
field must be dynamical.

In this paper we discuss the complete collection of currently
available observational and theoretical constraints on
unit-vector-tensor theories, combining our new result for the
remaining post-Newtonian parameters with previously established
constraints. Surprisingly, all of these constraints are compatible
with ranges of order unity for two coefficients in the Lagrangian.
We are aware of no other theory that comes this close to so many
predictions of general relativity and yet is fundamentally
different.

We call the timelike unit vector the aether, and the coupled
theory Einstein--aether theory, or ae-theory for short. A
review of its properties together with a collection of references
to earlier work was given in Ref.~\cite{dfest}. Ae-theory is
defined by the action for the metric $g_{ab}$ and aether $u^a$.
The most general action that is diffeomorphism-invariant and
quadratic in derivatives is
\begin{equation} \label{eq:action}
    S=\frac{-1}{16\pi G }
        \int \!d^4x \sqrt{-g} \Bigl(R+K^{ab}{}{}_{mn}
            \nabla_a u^m \nabla_b u^n +\lambda (u^a u_a -1)\Bigr)
\end{equation}
where
\begin{equation}\label{K}
    K^{ab}{}{}_{mn}=c_1\,  g^{ab} g_{mn} + c_2\,  \delta^a_m \delta^b_n
            + c_3\,  \delta^a_n \delta^b_m + c_4\,  u^a u^b g_{mn}.
\end{equation}
The coefficients $c_{1,2,3,4}$ are dimensionless constants, $R$ is
the Ricci scalar, and $\lambda$ is a Lagrange multiplier that
enforces the unit constraint. The expression $R_{ab} u^a u^b$ is
proportional to the difference of the $c_2$ and $c_3$ terms via
integration by parts, hence is not an independent term. Note that
since the covariant derivative of $u^a$ involves the Levi--Civita
connection, which involves first derivatives of the metric, the
aether part of the action in effect contributes also to the
metric kinetic terms. We adopt the metric signature
$({+}{-}{-}{-})$, and units are chosen such that the speed of
light defined by the metric $g_{ab}$ is unity. Other than the
signature choice we use the conventions of Ref.~\cite{wald}.

Observations have already severely constrained Lorentz symmetry
violation in the matter sector~\cite{Mattingly:2005re,Bluhm:2005uj},
hence to a very good approximation
matter must couple universally to one metric, which we take to be
$g_{ab}$. Our goal is to determine the observational and
theoretical constraints on the dimensionless parameters $c_i$.

In the weak-field, slow-motion limit ae-theory reduces to
Newtonian gravity~\cite{CLim}, with a value of Newton's constant
$G_{\rm N}$ related to the constant $G$ in (\ref{eq:action}) by
\beq\label{GNewton}
    G_{\rm N} = G\Big(1-\frac{c_{14}}{2}\Big)^{-1}.
\eeq
Here we have written $c_{14}$ for $c_1+c_4$, a notation
we generalize below to $c_{123}=c_1+c_2+c_3$, etc.

A standard way of beginning to compare an alternative gravity
theory to general relativity is to examine the first
post-Newtonian corrections. For a general metric theory of gravity
there are ten `parametrized post-Newtonian' (PPN)
parameters~\cite{WillBook,WillReview} characterizing the lowest
order effects in $v^2/c^2$ and dimensionless gravitational
potential $G_{\rm N}M/c^2r$. Five of these parameters,
$\zeta_1$,$\zeta_2$, $\zeta_3$, $\zeta_4$, and $\alpha_3$, vanish
identically for any `semi-conservative' theory, i.e. one derived,
like ae-theory, from a covariant action principle. Two others,
known as the Eddington--Robertson--Schiff parameters $\beta$ and
$\gamma$, characterize respectively the nonlinearity and the
spatial curvature produced by gravity. It was previously shown in
Ref.~\cite{ElingJ1} that in ae-theory $\beta=\gamma=1$, just as
in general relativity. Of the remaining three PPN parameters, two,
$\a_1$, $\a_2$, characterize preferred frame effects, and the
third, $\xi$ (sometimes called the Whitehead parameter),
characterizes a peculiar sort of three-body interaction.  The
parameter $\a_2$ for ae-theory was computed in
Ref.~\cite{Graesser:2005bg} to lowest nontrivial order in the
parameters $c_i$.

Here we report on an \textit{ab initio} computation of all the PPN
parameters that confirms the previous results and determines the
exact values of $\a_2$ and the previously undetermined parameters
$\a_1$ and $\xi$. The parameters are defined by a weak-field
expansion of the metric in terms of a set of generalized
potentials defined by integrals over a fluid source. To determine
the parameters, one solves the approximate field equations with
the fluid source in a standard coordinate gauge. The computation
is straightforward but lengthy, so the details are relegated to an
appendix.

Our results indicate that the ``time-time" and ``space-space"
components of the metric are the same in ae-theory and GR to
calculated post-Newtonian order, where we refer to a nearly
globally Lorentz coordinate system specialized to the standard PPN
gauge and with the aether aligned with the time direction at
zeroth-order. The ``time-space" components of the metric $g_{0i}$,
$i=1,2,3$, differ as
\beq
    (g_{0i})_{\rm ae} - (g_{0i})_{\rm GR}
        = \frac{\a_1-\a_2}{2}V_i + \frac{\a_2}{2}W_i,
\eeq
where $\a_{1,2}$ are the PPN parameters given explicitly below,
while the components of the aether are
\begin{gather}
    u^0 = 1 + U\\
    u^i = \frac{2c_1+3c_2+c_3+c_4}{2 c_{123}}\big(V_i - W_i\big)
            +\frac{(2-c_{14})(c_1-c_3)}{2c_1-c_1^2+c^2_3}\big(V_i+W_i\big).
\end{gather}
The potentials $U$, $V_i$, and $W_i$ are defined by
\beq
\begin{Bmatrix}
    U(t,\mathbf{x})\\V_i(t,\mathbf{x})\\W_i(t,\mathbf{x})\end{Bmatrix}
        = G_{\rm N}\int d^3 y\,\frac{\rho}{|\mathbf{x}-\textbf{y}|}
        \begin{Bmatrix}
            1\\v^i\\\frac{v^j (x^j-y^j)(x^i - y^i)}{|\mathbf{x}-\mathbf{y}|^2}\end{Bmatrix},
\eeq
where $\rho(t,\mathbf{y})$ is the rest-mass energy density of the
fluid source, $v^\mu(t,\mathbf{y})$ is the fluid four-velocity,
and the repeated indices are summed over.

The components of the perturbed
metric show that  the ae-theory PPN parameters are given by
\begin{gather}\label{PARA}
    \gamma = \beta = 1\\
    \xi = \zeta_1=\zeta_2=\zeta_3=\zeta_4=\alpha_3 = 0\\
    \alpha_1= \frac{-8(c_3^2 + c_1c_4)}{2c_1 - c_1^2+c_3^2}\\
    \alpha_2=\frac{(2c_{13}-c_{14})^2}
{c_{123}(2-c_{14})}\nonumber\\
-\frac{12c_3c_{13}+2c_1c_{14}(1-2c_{14})+
(c_1^2-c_3^2)(4-6c_{13}+7c_{14})} {(2-c_{14})(2c_1 - c_1^2+c_3^2)}.
\end{gather}
Note that the parameters $\a_1$ and $\a_2$ are both
of linear order in $c_i$ when the coefficients are small compared
to unity and the ratios amongst them are of order unity.

It is evident from the form of the metric and aether
perturbations that the cases $c_{123}=0$, $c_{14}=2$, and
$2c_1-c_1^2+c_3^2=0$ are special, since $\a_1$ and/or $\a_2$
diverges. Presumably the post-Newtonian approximation is not valid
when the coefficients are close to these values. From the wave
speeds (\ref{speeds}) below we see that the spin-0 speed vanishes
in either of the first two cases and the spin-1 speed vanishes in
the last case. This corresponds to the absence of spatial gradient
terms in the action~\cite{Graesser:2005bg}. The case $c_{123}=0$
corresponds~\cite{ElingJ1} to the vector-tensor theory of Hellings
and Nordtvedt~\cite{HellNord} if the unit constraint on the
aether is dropped.  This theory was shown by
Will~\cite{WillBook} to be dynamically over-determined and hence
observationally unacceptable.

The current best constraints~\cite{WillBook, WillReview} on the
preferred frame PPN parameters are $|\a_1|\lesssim 10^{-4}$ (from an
orbital polarization effect bounded by lunar laser ranging and
binary pulsar observations) and  $|\a_2|\lesssim 4\times10^{-7}$
(from a spin precession effect bounded by the alignment of the
solar spin with the ecliptic). These two conditions can be met with
two unrestricted parameters to spare, since we begin with
four free parameters $c_{1,2,3,4}$. At the lowest resolution,
we can just impose $\a_1=\a_2=0$.
The condition $\a_1=0$ implies
$c_4=-c_3^2/c_1$. (The case $c_1=0$
can be excluded since it yields $\a_1=-8$, which lies far outside the
observational bound.) Having put $\a_1=0$ in this way, it can
be seen (the seeing made easier with the help of Mathematica)
that $\a_2$ can be put to zero in two ways. One way is with
$c_{13}=0$, which together with the previous condition implies
also $c_{14}=0$.
This case is degenerate, and is
briefly discussed at the end of this paper.
The other way to put both $\a_1$ and $\a_2$ to zero is to determine
$c_2$ and $c_4$ in terms of $c_1$ and $c_3$ via
\beq\label{zeroalphas}
\begin{split}
    c_2&=(-2c_1^2-c_1c_3 + c_3^2)/3c_1\\
    c_4&=-c_3^2/c_1
\end{split}
\eeq
Thus {\it there is a two-parameter family of ae-theory Lagrangians
for which all the PPN parameters are identical to those of GR}.

We now consider the other constraints on ae-theory. In alternate
gravity theories including Brans--Dicke theory, the Newton
constant $G_{\rm N}$ need not be constant in time. Observational
bounds on $\dot{G}/G$ then constrain the theory. In the case of
ae-theory, we have the relation~\eqref{GNewton}, hence $G_{\rm
N}$ is always constant.

Another constraint arises from the possible
discrepancy between Newton's constant and the
gravitational constant occurring in the
equation for the dynamics of the cosmological scale factor.
In general relativity, the scale factor satisfies the
Friedman equation, which involves Newton's constant.
In ae-theory, when the metric has the standard cosmological
form (Robertson-Walker symmetry) and the aether
is  aligned with the cosmological rest frame,
the aether stress tensor can be constructed purely from the
spacetime metric with two derivatives,
and must be identically divergence free.
It must therefore be a linear combination of the Einstein
tensor $G_{ab}$ and a tensor constructed with the
spatial curvature scalar ${}^{(3)}R$, which turns out to be~\cite{MJindy,CLim}:
\beq\label{Taether}
    T^{\rm aether}_{ab}= -\frac{c_{13}+3c_2}{2}
        \Bigl[G_{ab}-\frac16{}^{(3)}R(g_{ab}+2u_au_b)\Bigr].
\eeq
The effect of the cosmological aether is thus to renormalize
the gravitational constant and to add a stress tensor of
perfect fluid type that in effect
renormalizes the spatial curvature contribution to the field
equations. The renormalized, cosmological gravitational constant
is given by~\cite{CLim}
\beq
    G_{\rm cosmo}=G\Big(1+\frac{c_{13}+3c_2}{2}\Big)^{-1}.
\eeq
Since this is not the same as $G_{\rm N}$ (\ref{GNewton}),
the expansion rate of
the universe differs from what would have been expected in GR with
the same matter content. The ratio is constrained by the observed
primordial ${}^4$He abundance to satisfy $|G_{\rm cosmo}/G_{\rm N}
- 1|<1/8$, which imposes a constraint on the constants
$c_i$~\cite{CLim}. Remarkably, if the constants are restricted by
(\ref{zeroalphas}) so that $\a_{1,2}=0$, then $G_{\rm N}=G_{\rm
cosmo}$. Primordial nucleosynthesis then imposes no additional
constraint.

Even when the two gravitational constants coincide, the
``curvature fluid" term in (\ref{Taether}) represents a deviation
from the Friedman equation in GR if the universe has non-zero
spatial curvature. Observations have shown that the spatial
curvature must be very small today, and it would have been even
less important in the past when the relative contribution of
matter and radiation would have been even more important. It thus
seems unlikely that an interesting constraint can be obtained from
this term. Another potential source of cosmological constraint is
the modification of the primordial fluctuation
spectrum~\cite{Lim}, but this has not yet been worked out in full
detail.

A further constraint on ae-theory comes from the possibility that
the gravity and aether waves travel at less than the speed of
``light"---that is, less than the limiting speed determined by the
metric $g_{ab}$ governing the propagation of matter fields. In
this case, high energy matter moving inertially through the vacuum
would produce vacuum \v{C}erenkov radiation of gravitational and
aether shock waves. A detailed analysis of this process and the
corresponding observational constraints from ultra-high-energy
cosmic ray observations was carried out in Ref.~\cite{Elliott:2005va}.
The constraints are characterized by very small numbers, ranging
between $10^{-15}$ and $10^{-31}$, depending on the wave-mode type
and emission process. These are all one-sided constraints, since
they apply only when the wave speeds are {\it smaller} than the
speed of light. To a first approximation then, the constraints
imply that the wave speeds must be greater than the speed of
light.

Some authors~\cite{CLim,Lim,Elliott:2005va}
have suggested that superluminal propagation be
excluded {\it a priori} on the grounds that ae-theory should be
viewed as an effective description of an underlying Lorentz
invariant theory in a configuration with broken Lorentz symmetry.
However, this is a purely theoretical bias, with no observational
basis that we can see. Moreover, superluminal propagation does not
threaten causality, as long as there is a limiting speed in at
least one given reference frame, as there is in ae-theory.  We
thus adopt a phenomenological stance, allowing for superluminal
propagation unless-and-until it is observationally ruled out.

There are five gravitational and aether wave modes in ae-theory:
two correspond to the usual gravitational spin-2 waves, two are a
transverse spin-1 aether-gravity wave, and one is a longitudinal
spin-0 aether-gravity wave. The squared speeds of these modes
are determined by the constants $c_i$, and are given by~\cite{JMwaves}
\beq\label{speeds}
\begin{split}
    \mbox{spin-2}\qquad&1/(1-c_{13})\\
    \mbox{spin-1}\qquad&(c_1-\half c_1^2
        +\half c_3^2)/c_{14}(1-c_{13})\\
    \mbox{spin-0}\qquad&c_{123}(2-c_{14})/c_{14}(1-c_{13})(2+c_{13}+3c_2).
\end{split}
\eeq
If we impose the $\a_{1,2}=0$ conditions (\ref{zeroalphas}), the
\v{C}erenkov constraint that the spin-2 and spin-0 wave speeds be
superluminal restricts $c_1$ and $c_3$ to the region
\beq\label{superluminal}
\begin{split}
    &0<c_{13}<1\\
    &0<(c_1-c_3)< c_{13}/3(1-c_{13}).
\end{split}
\eeq
These conditions also ensure that the spin-1 wave speed is
superluminal.

In addition to the observational constraints already mentioned,
there are two theoretical constraints, coming from the requirement
that the wave modes be stable---i.e.~have real frequencies---and
that the energy of the modes be positive. The first requirement is
already guaranteed by the no-\v{C}erenkov condition that the
speeds be greater than unity. The signs of the energy densities of
the wave modes, averaged over a cycle, are given
by~\cite{Eling:2005zq}
\beq\label{energy}
\begin{split}
    \mbox{spin-2}\qquad&1\\
    \mbox{spin-1}\qquad&(2c_1-c_1^2+c_3^2)/(1-c_{13}) \\
    \mbox{spin-0}\qquad&c_{14}(2-c_{14}).
\end{split}
\eeq
The spin-2 modes always carry positive energy. If the $\a_{1,2}=0$
conditions (\ref{zeroalphas}) and the superluminal conditions
(\ref{superluminal}) are satisfied, then we find that the spin-1
and spin-0 modes also carry positive energy. By contrast, if the
speeds are sub-luminal, then the latter two modes carry negative
energy. Thus, not only the \v{C}erenkov constraints, but also the
positive energy requirement excludes the case of sub-luminal wave
speeds.

We earlier pointed out that an alternate way to set $\a_1=\a_2=0$
is if $c_{13}=c_{14}=0$. In this case $G_{\rm N}/G_{\rm cosmo}=(1
+3c_2/2)$, so nucleosynthesis would impose a constraint on $c_2$.
A more serious concern is that the spin-0 and spin-1 wave speeds
(\ref{speeds}) diverge in this case, because there are no time
derivative terms in the aether field equation~\cite{JMwaves}.
At the same time the energy density (\ref{energy}) of the spin-0
mode goes to zero, but that of the spin-1 mode remains finite. We
have not worked out the observational signatures of this behavior.

It is rather non-trivial that the PPN parameters are {\it
identical} to those of GR and that the vacuum \v{C}erenkov,
nucleosynthesis, stability, and positive energy constraints are
all satisfied in a large two-dimensional region
(\ref{zeroalphas},\ref{superluminal}) in the four-dimensional
$c_i$ parameter space. To further constrain the parameters one
should look to strong-field effects or radiative processes. Since
the gravitational radiation damping of binary pulsars has been
found to agree with GR to one part in $10^3$, one would expect
constraints of order $10^{-3}$ from such observations. The
ae-theory radiation damping rate has been determined to lowest
non-trivial order~\cite{AERAD}, but higher-order terms
are still needed to set definitive constraints.  

One particular strong-field effect that can be examined is the
existence and nature of black hole solutions to the vacuum field
equations. Some alternate theories of gravity whose PPN parameters
are equal or close to those of GR do not admit regular black hole
solutions. In these theories, astrophysical collapse would produce
something other than a black hole---perhaps a naked singularity,
or a bounce---which may not be difficult to rule out
observationally. Recent studies~\cite{ElingJbh} have shown,
however, that ae-theory does admit regular black hole solutions.

In conclusion, ae-theory provides a healthy sparring partner for
GR, a role that was previously played only by scalar-tensor
theory. One of the motivations for exploring this theory is the
idea that the existence of a preferred timelike direction in
spacetime could play a role in the solution to fundamental
problems such as the nature of dark energy or quantum gravity.
Although it is likely that radiation and strong-field constraints
will limit all of the theory's parameters to be small compared to
unity, they would still allow the  aether field to {\it exist}
and possibly play such a role.
%
%
\begin{acknowledgments}
We thank Konstantin Zloshchastiev for noting
several typos in an earlier draft of the appendix.
This research was supported in part by the NSF under grant
PHY-0300710 at the University of Maryland.
\end{acknowledgments}
%
%
\appendix*
\section{Calculation of ae-theory PPN parameters}

This appendix provides details of the calculation of the
Parameterized Post-Newtonian (PPN) parameters,
$\alpha_1$,$\alpha_2$,$\alpha_3$,$\beta$,$\gamma$,$\zeta_1$,$\zeta_2$,$\zeta_3$,$\zeta_4$,$\xi$,
for ae-theory.  The PPN formalism is defined in a weak-field,
slow-motion limit, and describes the next-to-Newtonian order
gravitational effects in terms of a standardized set of potentials
and these ten parameters. We will determine the PPN parameters by
solving the field equations with a perfect-fluid source in a
standard coordinate gauge order-by-post-Newtonian-order.  More
detailed explanations of the procedure and the general PPN
formalism can be found in the classic reference of
Will~\cite{WillBook}.

We will follow the conventions of~\cite{wald}, with the following
exceptions and additions.  The metric will have signature 
$({+},{-},{-},{-})$.
We will write the approximate equations in terms of components of
tensors with all indices lowered. Spatial indices will be
indicated by lowercase Latin letters from the middle of the
alphabet: ${i,j,k,\dots}$.  Spatial indices are raised and lowered
with the spacetime metric; e.g.~$v_i \equiv v^a g_{ai} = -v^i +
(\text{higher order terms})$. Repeated spatial indices are summed
over.  We will write, e.g.~$c_{14}$ for $c_1 + c_4$, etc.  Note also 
that we follow modern conventions for denoting the post-Newtonian
order of quantities (see, e.g.~\cite{WillReview}), which differ from the 
conventions of~\cite{WillBook}.

The ae-theory field equations follow from the
action~\eqref{eq:action}, with an additional perfect-fluid source
coupled in the standard way to the metric $g_{ab}$, and uncoupled
to the aether $u^a$. We have the Einstein equations, written in
the non-standard form
\beq\label{EEQ}
    R_{ab} = \big(S_{cd} + 8\pi G T_{cd}\big)
            \big(\delta^c_a \delta^d_b - \half g_{ab}g^{cd}\big),
\eeq
where
\beq\begin{split}
    S_{ab} = &\nabla_m\bigl(J_{(a}^{\ph{(a}m}u_{b)} -
        J^m_{\ph{m}(a} u_{b)} - J_{(ab)}u^m\bigr)\\&
        +c_1\bigl(\nabla_m u_a\nabla^mu_b
            - \nabla_a u_m\nabla_b u^m\bigr)
        +c_4{\dot u}_a {\dot u}_b\\
        &+\lambda u_a u_b +\frac{1}{2}g_{ab}
            (J^n_{\ph{a}m}\nabla_n u^m);
\end{split}\eeq
with
\beq
    J^a_{\phantom{a}m} = \big(c_1g^{ab}g_{mn}
        +c_2\delta^a_m\delta^b_n + c_3\delta^a_n\delta^b_m
        +c_4 u^a u^b g_{mn}\big)
        \nabla_b u^n,
\eeq
and
\beq
    {\dot u}^a = u^b\nabla_b u^a;
\eeq
also
\beq
    T^{ab} = (\rho + \rho\Pi + p)v^a v^b - p g^{ab},
\eeq
where $v^a$ is the four-velocity, $\rho$ the rest-mass-energy
density, $\Pi$ the internal energy density, and $p$ the isotropic
pressure of the fluid.  We also have the aether field equation
\beq\label{AEQ}
    \nabla_a J^a_{\phantom{a}m}-c_4 {\dot u}_a \nabla_m u^a
        = \lambda u_m;
\eeq
and the constraint
\beq\label{CON}
    g_{ab}u^a u^b = 1.
\eeq
Eqn.~\eqref{AEQ} can be used to eliminate $\lambda$, giving
\beq
    \lambda = u^m\nabla_a J^a_{\phantom{a}m}
    - c_4{\dot u}^a{\dot u}_a.
\eeq

We assume a nearly-globally-Lorentzian coordinate system and basis
with respect to which, at zeroth order, the metric is the
Minkowski metric $\eta_{ab}$ and the aether is purely
timelike. The fluid variables are assigned orders of $\rho \sim
\Pi\sim p/\rho\sim (v^i)^2\sim O(1)$. Taking the time-derivative
of a quantity will effectively raise its order by one-half:
$X \sim O(N) \rightarrow \partial X/\partial t \sim O(N+1/2)$. 
We assume
that the components of the metric perturbations 
$h_{ab}$ with respect to this basis will
be of orders
\beq
    h_{00} \sim O(1) + O(2),\quad h_{ij} \sim O(1),\quad h_{0i} \sim
    O(1.5).
\eeq
This assignment preserves the Newtonian limit while allowing one
to determine just the first post-Newtonian corrections. The
aether perturbations $\delta u^a$ are assumed to be of orders
\beq
    \delta u^0 \sim O(1),\qquad \delta u^i \sim O(1.5).
\eeq
Lower orders are disallowed by the field equations, given the
above orders of $h_{ab}$.  We will assume that $h_{ab}$
and $\delta u^a$ satisfy boundary conditions such that they vanish at
spatial infinity.

The metric components are to be expanded in terms of particular
potential functions, thus defining the PPN parameters:
\beq\label{STNDRD}
\begin{split}
    g_{00} &= 1-2U+2\beta
        U^2+2\xi\Phi_W-(2\gamma+2+\alpha_3+\zeta_1-2\xi)\Phi_1\\
        &\quad-2(3\gamma-2\beta+1+\zeta_2+\xi)\Phi_2 
        -2(1+\zeta_3)\Phi_3-2(3\gamma+3\zeta_4-2\xi)\Phi_4
	+(\zeta_1-2\zeta)\mathcal{A},\\
    g_{ij}&=-(1+2\gamma U)\delta_{ij},\\
    g_{0i}&=\half(4\gamma+3+\alpha_1-\alpha_2+\zeta_1-2\xi)V_i
        +\half(1+\alpha_2+\zeta_1+2\xi)W_i.
\end{split}\eeq
The potentials are all of the form
\beq
    F(x) =G_{\rm N} \int d^3y \frac{\rho(y)f}{|x-y|},
\eeq
where $G_{\rm N}$ is the current value of Newton's constant, which
we determine below in terms of $G$ and the $c_i$. The
correspondences $F:f$ are given by
\begin{gather}\label{POTS}
    U:1\qquad
    \Phi_1 : v_i v_i\qquad
    \Phi_2 : U\qquad
    \Phi_3 : \Pi\qquad
    \Phi_4 : p/\rho\nonumber\\
    \Phi_W : \int d^3z\;
    \rho(z)\frac{(x-y)_j}{|x-y|^2}
    \Big(\frac{(y-z)_j}{|x-z|}-\frac{(x-z)_j}{|y-z|}\Big)\qquad
    \mathcal{A} : \frac{(v_i (x-y)_i)^2}{|x-y|^2}\\
    V_i : v^i\qquad
    W_i : \frac{v_j (x_j - y_j)(x^i -y^i)}{|x-y|^2}\nonumber.
\end{gather}
Note that for $U$, $\Phi_{1,2,3,4}$, and $V_i$,
\beq
    F_{,ii} = -4 \pi G_{\rm N}\, \rho f.
\eeq
We will also make use of the `superpotential' $\chi$:
\beq \chi = -G_{\rm N} \int d^3y\, \rho |x-y|, \eeq
which satisfies
\beq\label{chi}
    \chi_{,ii} = -2 U.
\eeq
We also note the relation
\begin{gather}\label{XVW}
    \chi_{,i0} = V_i - W_i,
\end{gather}
which follows from the formula
\beq
    \frac{\partial}{\partial t}\int d^3y\,
        \rho(\mathbf{y},t) f(\mathbf{x,y})
        = \int d^3y\,\rho(\mathbf{y},t)
            v^i(\mathbf{y},t)\frac{\partial f}{\partial y^i}
                [1+O(1)],
\eeq
which follows from the continuity equation for the fluid
\beq
    \rho_{,0} + (\rho v^i)_{,i} = 0,
\eeq
assumed to hold to $O(1.5)$.

 These potentials satisfy certain criteria of
``reasonableness" and simplicity (see~\cite{WillBook}, Sec. (4.1)
for details), and are general enough to describe all known viable
theories of gravity.  In particular, they suffice for ae-theory.
The criteria permit $g_{00}$ to depend also on the potential
$\chi_{,00}$, and $g_{ij}$ to depend on $\chi_{,ij}$. Such terms,
however, can always be eliminated~\cite{WillBook} by a suitable
coordinate transformation that preserves the zeroth-order form of
the components. The `standard PPN gauge' is thus defined as that
post-Newtonian coordinate frame in which all dependence on
$\chi_{,00}$ and $\chi_{,ij}$ has been removed from, respectively,
$g_{00}$ and $g_{ij}$. This fixing determines the coordinate frame
up to necessary order so that the standard form of the metric
components is unambiguous.

In carrying out the calculations, we shall impose the following
gauge conditions:
\begin{gather}
    h_{ij,j} = -\half (h_{00,i}-h_{jj,i})\label{GAG1}\\
    h_{0i,i} = -3 U,_0+ \theta n_{i,i}\label{GAG2}
\end{gather}
where $\theta$ is an arbitrary parameter and $n_i = u_i - h_{0i}$.
These conditions are suggested by the standard conditions for  
general relativity.  
As we shall see, the conditions~\eqref{GAG1} suffice to put
$g_{ij}$ in standard form, while the fourth condition~\eqref{GAG2} 
standardizes
$g_{00}$ when
\beq
    \theta = -\frac{(c_1+2c_3-c_4)}{(2 - c_{14})}.
\eeq

The solving procedure is as follows:
\begin{verse}
Step 1: Solve the constraint~\eqref{CON} for $u^0$ to $O(1)$; \\
Step 2:    Solve the ``time-time" component of the Einstein
equation~\eqref{EEQ} for
$g_{00}$ to $O(1)$; \\
 Step 3:   Solve the ``space-space" components
of~\eqref{EEQ} for $g_{ij}$ to $O(1)$;\\
Step 4:    Solve the ``space" components of the aether field
equation~\eqref{AEQ} for $u^i$ to
$O(1.5)$; \\
  Step 5:  Solve the ``time-space" components of~\eqref{EEQ} for
$g_{0i}$ to $O(1.5)$; \\
 Step 6:   Solve the ``time-time" component
of~\eqref{EEQ} for $g_{00}$ to $O(2)$.
\end{verse}
The cases in which $c_{123} = 0$, $c_{14} = 2$, or $2c_1 -
c_1^2+c_3^2 = 0$ are special in that the found solutions diverge.
We will presume that the post-Newtonian approximation is not valid
in these cases, and assume below that they do not hold.  See the
main text for more discussion of this point.
%
%
%
\subsection{$u^0$ to $O(1)$}
Solving the constraint~\eqref{CON} gives
\beq\label{AEO}
    u^0 = 1 - (1/2)h_{00}
\eeq
to $O(1)$.  For the components of $u_a$, we have
\beq\label{UO}
    u_0 \equiv u^a g_{a 0} = 1 + \half h_{00},
\eeq
and
\beq
     u_i = u^a g_{a i}= -u^i + h_{0i} = n_i+h_{0i}.
\eeq

For later convenience, we will now express the covariant
derivatives of $u_a$.  The constraint~\eqref{CON} implies that
\beq
    \nabla_a u_0 = 0
\eeq
to $O(2)$.  Also to $O(2)$, we have
\beq
    \nabla_0 u_i = -\half h_{00,i}(1-\half h_{00}) + h_{0i,0} + n_{i,0},
\eeq
and
\beq
    {\dot u}_i = u^0 \nabla_0 u_i = -\half h_{00,i}(1-h_{00})
                                       + h_{0i,0}+ n_{i,0}.
\eeq
To $O(1.5)$, we have
\beq
    \nabla_j u_i = n_{i,j} + \half h_{ij,0} + h_{0[i,j]}.
\eeq
%
%
%
\subsection{$g_{00}$ to $O(1)$}

We now solve the ``time-time" component of the Einstein
equation~\eqref{EEQ} for $g_{00}$ to $O(1)$. For the components of
$R_{00}$, we have
\beq\label{ROOa}
    R_{00} = \half h_{00,ii}
    + \half h_{ij}h_{00,ij} -
    \big(h_{i0,i} - \half h_{ii,0}\big)_{,0}
    -\frac{1}{4} h_{00,i}h_{00,i}
    +\frac{1}{4} h_{00,j}\big(2h_{ij,i} - h_{ii,j}\big)
\eeq
to $O(2)$. At $O(1)$, we have
\begin{gather}
    R_{00}=\half h_{00,ii},\\
    T_{00}=\rho,\qquad
    T_{ij} = 0,\label{TOO}\\
    S_{00}= J_{0\ph{m},m}^{\ph0 m} - J_{00,0} = -J_{0i,i}
    = -c_{14} (\nabla_0 u_i)_{,i} =
    \frac{c_{14}}{2}h_{00,ii},\label{SOO}\\
    S_{ij}= 0\label{SIJ}.
\end{gather}
The field equation becomes
\beq
    (1-\frac{c_{14}}{2})h_{00,ii} = 8\pi G \rho,
\eeq
which gives $h_{00}$ to O(1),
\beq\label{HOO}
    h_{00} = -2 U,
\eeq
with Newton's constant
\beq\label{GNewt}
    G_N = \big(1-\frac{c_{14}}{2}\big)^{-1} G.
\eeq
%
%
\subsection{$g_{ij}$ to $O(1)$}

We now solve the ``space-space" components of~\eqref{EEQ} for
$g_{ij}$ to $O(1)$. We have to $O(1)$
\beq\begin{split}
    R_{ij} &= \half h_{ij,kk}+\half h_{kk,ij}-h_{k(i,j)k}
        -\half h_{00,ij}\\
        &=\half h_{ij,kk},
\end{split}\eeq
where we have imposed the gauge condition~\eqref{GAG1} in the
second step.  Using~\eqref{TOO},~\eqref{SOO}, and~\eqref{SIJ}, the
field equation becomes
\beq
    h_{ij,kk} = 8\pi G_N \rho\, \delta_{ij},
\eeq
giving
\beq\label{HIJ}
    h_{ij} = -2 U \delta_{ij}.
\eeq
%
%
%
\subsection{$u^i$ to $O(1.5)$}

We now solve the ``space" components of the aether field
equation~\eqref{AEQ} for $u^i$ to $O(1.5)$, making use of the gauge
condition~\eqref{GAG1} and our earlier
results~\eqref{UO},~\eqref{HOO}, and~\eqref{HIJ}. At $O(1.5)$ 
equation~\eqref{AEQ} has the form
\beq
    J^a_{\ph a i,a} = J_{0i,0}-J_{ji,j}= 0.
\eeq
To $O(1.5)$,
\beq\label{JOI}
    J_{0i,0} = c_{14} (\nabla_0 u_i)_{,0} =
    -\frac{c_{14}}{2}h_{00,i0}= -\frac{c_{14}}{2}\chi_{,0ijj},
\eeq
and
\beq\begin{split}
    J_{ji,j} &=\big(c_1\nabla_j u_i + c_2 \delta_{ij} \nabla_k u_k
    + c_3 \nabla_i u_j\big)\\
        &=c_1 n_{i,jj}+c_{23}n_{j,ji}
            +\half\big(2c_-h_{0[i,j]j}+(c_++3c_2)\chi_{,0ijj}\big),
\end{split}
\eeq
where $c_- = c_1 - c_3$, $c_+ = c_{13}$.
The aether field equation can then be written
\beq\label{AEQI}
    \Big(c_1 n_i + \frac{c_-}{2}h_{0i}
        +\half(2c_1 + 3c_2 + c_3+c_4)\chi_{,i0}\Big)_{,jj}
        -\big(\frac{c_-}{2}h_{0j,j} -c_{23}n_{j,j}\big)_{,i}
        =0.
\eeq
Taking the spatial divergence of the left-hand side gives the
relation
\beq
    n_{i,ijj} = A \chi_{,0iijj},
\eeq
where
\beq\label{AAA}
    A = -\frac{2c_1 + 3c_2 + c_3 + c_4}{2 c_{123}},
\eeq
which we can solve for $n_{i,i}$. Substituting into~\eqref{AEQI},
imposing the gauge condition~\eqref{GAG2}, and using our earlier
results, we solve~\eqref{AEQI}:
\beq\label{AEI}
    n_i = -u^i=-\frac{1}{2c_1}\Big(c_- h_{0i}
    -\big(2c_1 A + c_-(\frac{3}{2}+A\theta)\big)
    \chi_{,0i}\Big).
\eeq

%
%
\subsection{$g_{0i}$ to $O(1.5)$}

We now solve the ``time-space" components of~\eqref{EEQ} for
$g_{0i}$ to $O(1.5)$, making use of the gauge
conditions~\eqref{GAG1} and~\eqref{GAG2} and the earlier
results~\eqref{UO},~\eqref{HOO},~\eqref{HIJ}, and~\eqref{AEI}.  We
have to $O(1.5)$
\beq\begin{split}
    R_{0i} &= h_{0[i,j]j} + h_{j[j,i]0} \\
        &=\half \big(h_{0i}
        +\half\big(1-2A\theta\big)
    \chi_{,0i}\big)_{,jj},
\end{split}\eeq
Also to $O(1.5)$,
\beq
    T_{0i}= \rho v_i,
\eeq
and
\beq
    S_{0i} = -J_{(i0),0} +\half J_{i\ph m,m}^{\ph i m}
        = -\half\big(J_{0i,0}+J_{ij,j}\big).
\eeq %
We have
\beq\begin{split}
    J_{ij,j}
    &= (c_1 \nabla_i u_j +c_2\delta_{ij}\nabla_k u_k+c_3\nabla_j u_i)_{,j}\\
        &=(c_{12}n_{j,i}+c_3n_{i,j})_{,j}
            +\half(2c_-h_{0[j,i]j}+c_+h_{ij,j0}+c_2h_{jj,i0})\\
    &= \Big(-\frac{c_- c_+}{2c_1}h_{0i}
            +(\frac{c_{14}}{2}-E) \chi_{,0i}\Big)_{,jj},
\end{split}
\eeq
where
\beq
    E = \frac{1}{4c_1}\Big(c_1^2 + 3c_3^2 + 4c_1 c_4
        - 2c_- c_+ A \theta\Big).
\eeq
With~\eqref{JOI}, this gives
\beq
    S_{0i} = \Big(\frac{c_- c_+}{4c_1}h_{0i} +
    \frac{E}{2}\chi_{,0i}\Big)_{,jj}.
\eeq
The field equation becomes
\beq
    (1-\frac{c_- c_+}{2c_1})h_{0i,jj} = 16\pi G \rho v_i
        +(E+A\theta-\half)\chi_{,0ijj},
\eeq
giving
\beq\label{HOI}
    h_{0i} = (1-\frac{c_-c_+}{2c_1})^{-1}
        \Big((E+A\theta-\half)\chi_{,0i}
        +4(1-\frac{c_{14}}{2})V_i\Big).
\eeq
%
%
%
\subsection{$g_{00}$ to $O(2)$}

We now solve the ``time-time" component of~\eqref{EEQ} for
$g_{00}$ to $O(2)$, making use of the gauge
conditions~\eqref{GAG1} and~\eqref{GAG2} and the earlier
results~\eqref{UO},~\eqref{HOO},~\eqref{HIJ},~\eqref{AEI},
and~\eqref{HOI}.  Define ${\tilde h}_{00} = g_{00} -1 + 2U$. Then,
from eqn.~\eqref{ROOa}, we have
\beq\label{ROOb}
    R_{00} = \half \big({\tilde h}_{00} - 2U
    -2U^2+8\Phi_2-2A\theta\chi_{,00}\big)_{,ii}.
\eeq
Also,
\beq
    T_{00} =\rho (1+\Pi+v_iv_i - 2U),
\eeq
\beq
     T_{ij} = \rho v_i v_j + p \delta_{ij}.
\eeq
\beq
    g_{00}(T_{ab}g^{ab})=(1-2U)(T^{00}(1-2U)-T^{ii}(1+2U)),
\eeq
so that
\beq\label{TOOb}\begin{split}
    T_{00}-\half g_{00}(T_{ab}g^{ab})
        &= \half \rho(1+\Pi+2(v_iv_i - U))+\frac{3}{2}p\\
        &=\frac{-\big(1-c_{14}/2\big)}{8 \pi G}
            \big(U+2\Phi_1 - 2\Phi_2+\Phi_3+3\Phi_4\big)_{,ii}.
\end{split}
\eeq
To aid the reader in sorting through the terms appearing in
$S_{ab}$, we note that to $O(2)$, in the chosen gauge,
\beq
    \nabla_i u_i =(\frac{3}{2}+A) \chi_{,0ii}
\eeq
and
\beq
    (\nabla_0 u_i)_{,i} = -\half\Big(-2U + {\tilde h}_{00}
        -U^2 -2\big(\frac{3}{2}+(1+\theta)A\big) \chi_{,00}\Big).
\eeq
After some bookkeeping, we find that
\beq
    S_{00}= \frac{c_{14}}{2}\big(-2U + {\tilde h}_{00}
        -\frac{5}{2}U^2+9\Phi_2\big)_{,ii}
        -c_{14}(\frac{3}{2}+(1+\theta)A)\chi_{,00ii},
\eeq
and
\beq
    S_{ii} = \frac{c_{14}}{2}\big(\half U^2 - \Phi_2)_{,ii}
    -(c_+ + 3c_2)(\frac{3}{2}+A)\chi_{,00ii}.
\eeq
We thus have
\beq\label{KOOb}\begin{split}
    S_{00} - \half g_{00}S_{ab}g^{ab} &=
    \half(S_{00} + S_{ii}) + 2US_{ii}\\
        &=\frac{c_{14}}{4}\Big(-2U +
        {\tilde h}_{00} -2 U^2 + 8 \Phi_2\Big)_{,ii}\\
        &\quad
        -\half\Big((\frac{3}{2}+A)(2c_1+3c_2+c_3+c_4)
         + c_{14}A\theta\Big)
        \chi_{,00ii}.
\end{split}
\eeq
We combine~\eqref{ROOb},\eqref{TOOb}, and~\eqref{KOOb}, solve the
field equation, and obtain
%
%
%
\beq\label{HOOb}
    {\tilde h}_{00}
    = 2U^2-4\Phi_1 -4\Phi_2 -2\Phi_3-6\Phi_4
        +Q\chi_{,00},
\eeq
where
\beq
    Q=(1-\frac{c_{14}}{2})^{-1}\big((2-c_{14})\theta+(c_1+2c_3-c_4)\big)A.
\eeq
Finally, we move into the standard gauge by choosing $\theta$ so
that $Q$ vanishes:
\beq\label{theta}
    \theta \longrightarrow \theta_0=-\frac{c_1+2c_3-c_4}{2-c_{14}}.
\eeq
%
%
%
\subsection{Summary}
We now collect our results,
eqns.~\eqref{HOO},~\eqref{HIJ},~\eqref{HOI}, and~\eqref{HOOb} for
the metric components, and~\eqref{AEO} and~\eqref{AEI} for the
aether, imposing the gauge conditions~\eqref{GAG1}
and~\eqref{GAG2} with $\theta=\theta_0$, and using the
relation~\eqref{XVW}. For the metric, we have
\begin{align}
    g_{00} &=1-2U+2U^2-4\Phi_1 -4\Phi_2 -2\Phi_3-6\Phi_4\\
    g_{ij} &=-(1+2U)\delta_{ij}\\
    g_{0i} &= \frac{2 c_1}{2c_1 -c_1^2+c_3^2}\Big(
        \big(E+A\theta_0-\half+2(2-c_{14})\big)V_i
        -\big(E+A\theta_0-\half\big)W_i\Big).
\end{align}
We can extract the PPN parameters by comparison with the standard
forms~\eqref{STNDRD}. We find
\begin{gather}
    \gamma = \beta = 1\\
    \xi = \zeta_1=\zeta_2=\zeta_3=\zeta_4=\alpha_3 = 0\\
    \alpha_1= -\frac{8(c_3^2 + c_1c_4)}{2c_1 - c_1^2+c_3^2}\\
\begin{split}
    \alpha_2&= \frac{(2c_{13}-c_{14})^2}{c_{123}(2-c_{14})}\\
    &\quad
        -\frac{12c_3c_{13}+2c_1c_{14}(1-2c_{14})+
        (c_1^2-c_3^2)(4-6c_{13}+7c_{14})} {(2-c_{14})(2c_1 -
        c_1^2+c_3^2)}.
\end{split}
\end{gather}
For the aether, we have
\begin{align}
    u^0 &= 1 + U\\
    u^i & = -\big(A+B\big)V_i + \big(A-B) W_i,
\end{align}
where
\beq
    A = -\frac{2c_1+3c_2+c_3+c_4}{2c_{123}},\qquad
    B = -\frac{(2-c_{14})c_-}{2c_1-c_1^2+c_3^2}.
\eeq
%
%

%
%

\begin{thebibliography}{99}
%
\bibitem{WillBook}C.M. Will, {\it Theory and Experiment in Gravitational
Physics}, (Cambridge Univ. Press, Cambridge, 1993).

\bibitem{WillReview}
C.~M.~Will,
``The confrontation between general relativity and experiment,''
Living Rev.\ Rel.\  {\bf 4}, 4 (2001)
[arXiv:gr-qc/0103036].

\bibitem{dfest}
  C.~Eling, T.~Jacobson and D.~Mattingly,
  ``Einstein--aether theory,''
  arXiv:gr-qc/0410001.

\bibitem{wald}R.~M.~Wald, {\it General Relativity}
(University of Chicago Press, 1984).

\bibitem{Mattingly:2005re}
  D.~Mattingly,
  ``Modern tests of Lorentz invariance,''
  arXiv:gr-qc/0502097.

\bibitem{Bluhm:2005uj}
  R.~Bluhm,
  ``Overview of the SME: Implications and phenomenology of Lorentz violation,''
  arXiv:hep-ph/0506054.

\bibitem{CLim}
S.~M.~Carroll and E.~A.~Lim,
``Lorentz-violating vector fields slow the universe down,''
Phys.\ Rev.\ D {\bf 70}, 123525 (2004)
[arXiv:hep-th/0407149].

\bibitem{ElingJ1}
C.~Eling and T.~Jacobson,
``Static post-Newtonian equivalence of GR and gravity with a dynamical
preferred frame,''
Phys.\ Rev.\ D {\bf 69}, 064005 (2004)
[arXiv:gr-qc/0310044].

\bibitem{Graesser:2005bg}
  M.~L.~Graesser, A.~Jenkins and M.~B.~Wise,
  ``Spontaneous Lorentz violation and the long-range gravitational
  preferred-frame effect,''
  Phys.\ Lett.\ B {\bf 613}, 5 (2005)
  [arXiv:hep-th/0501223].

\bibitem{HellNord}
    R.~W.~Hellings and K.~Nordtvedt, Jr.,
    ``Vector-Metric theory of
    gravity,"
    Phys.\ Rev.\ D {\bf 7}, 3593 (1973).

\bibitem{MJindy}
D.~Mattingly and T.~Jacobson,
``Relativistic gravity with a dynamical preferred frame,''
in {\it CPT and Lorentz
Symmetry II}, ed. V.A. Kostelecky (World Scientific, Singapore,
2002) [arXiv:gr-qc/0112012].

\bibitem{Lim}
E.~A.~Lim,
``Can We See Lorentz-Violating Vector Fields in the CMB?,''
arXiv:astro-ph/0407437.

\bibitem{Elliott:2005va}
  J.~W.~Elliott, G.~D.~Moore and H.~Stoica,
  ``Constraining the new aether: Gravitational Cherenkov radiation,''
  JHEP {\bf 0508}, 066 (2005)
  [arXiv:hep-ph/0505211].

\bibitem{JMwaves}
T.~Jacobson and D.~Mattingly, ``Einstein--Aether waves,'' Phys.\
Rev.\ D {\bf 70}, 024003 (2004) [arXiv:gr-qc/0402005].

\bibitem{Eling:2005zq}
  C.~Eling,
  ``Energy in the Einstein--aether theory,''
  arXiv:gr-qc/0507059.
\bibitem{AERAD}
  B.~Z.~Foster,
  ``Radiation damping in Einstein-aether theory,''
  arXiv:gr-qc/0602004.
\bibitem{ElingJbh}
C.~Eling and T.~Jacobson, in preparation.
%
%
\end{thebibliography}
\end{document}